%% file: sphericalChargeWithBBL.tex
\documentclass[aps,pre,twocolumn,showpacs,superscriptaddress,groupedaddress,amsmath]{revtex4}  
\usepackage{graphicx}  
\usepackage{dcolumn}   
\usepackage{bm}        
\usepackage{amssymb}   

\begin{document}
\renewcommand{\vec}[1]{\mathbf{#1}}
\widetext


\title{Dynamics of spherical distributions of charge\\with small internal dipolar motion}
\input author_list.tex       
\date{\today}

\begin{abstract}
This paper extends the Lorentz-Abraham model of an electron (i.e. the equations of motion for a small spherical shell of charge, which is rigid in its proper frame) to treat a small spherically symmetric charge distribution, allowing for small internal dipolar motion. This is done by dividing the distribution into thin spherical shells (in the continuum limit), and tracking the interactions between shells. Dipolar motion of each constituent spherical shell is allowed along the net dipole moment, but higher order multipole-moments are ignored. The amplitude of dipolar motion of each spherical shell is assumed to be linearly proportional to the net dipole moment. Under these assumptions, low velocity equations of motion are determined for both the center-of-mass motion and net dipolar motion of the distribution. This is then generalized to arbitrary (relativistic) center-of-mass velocity and acceleration, assuming the motion of individual shells is completely in phase or out of phase with the net dipole moment.
\end{abstract}

\maketitle

\section{Introduction}
The classical dynamics of a small charged spherical shell has been extensively studied for more than 100 years. The equation of motion, or the Lorentz-Abraham equation, is interesting due to the self-electromagnetic force on the shell, which results in radiative damping and momentum/energy transfer between the charge and its velocity fields\cite{lorentz1892,larmor1897,heaviside1902,abraham1904,abraham1905,vonlaue1909,schott1912,poincare1906,dirac1938}. Various aspects of the theory, such as apparent paradoxes (apparent discrepancy between force and power equations, the ``4/3 problem'', runaway self-acceleration solutions, and pre-acceleration) continue to be discussed in the literature\cite{jackson,medina2006,rohrlich2008,aguirregabiria2006,essen2015,ferris2011,heras2003,rohrlich2000,rohrlich2001,steane2015,steane2015b,steane2015c,villarroel2002}.  See Ref.~\cite{rohrlich1997} for a brief historical overview of the problem. A full history and detailed treatment of the spherical shell, with a description of the cause and resolution of the paradoxes may be found in the excellent monograph by Arthur Yaghjian~\cite{yaghjian}.

This paper treats the classical dynamics of any small spherical distribution of charge (to zero order in its size), built up of spherical shells in the continuum limit. These spherical shells are held concentric (and kept from exploding) by some binding force. When the distribution is accelerated, the spherical shells may displace from equilibrium, but if the distribution is stable, the binding force will return the spherical shells to concentricity (and likely create oscillations). The deviation of the binding force from its equilibrium value, which we will call the restoring force, will be approximated as linear (the lowest order term in a power series for the restoring force about equilibrium). I treat the non-relativistic (low-velocity) case first and then the arbitrary velocity case, maintaining a low-velocity assumption on the motion of individual shells about the center of mass.

\begin{figure}
{
\includegraphics[scale=0.4]{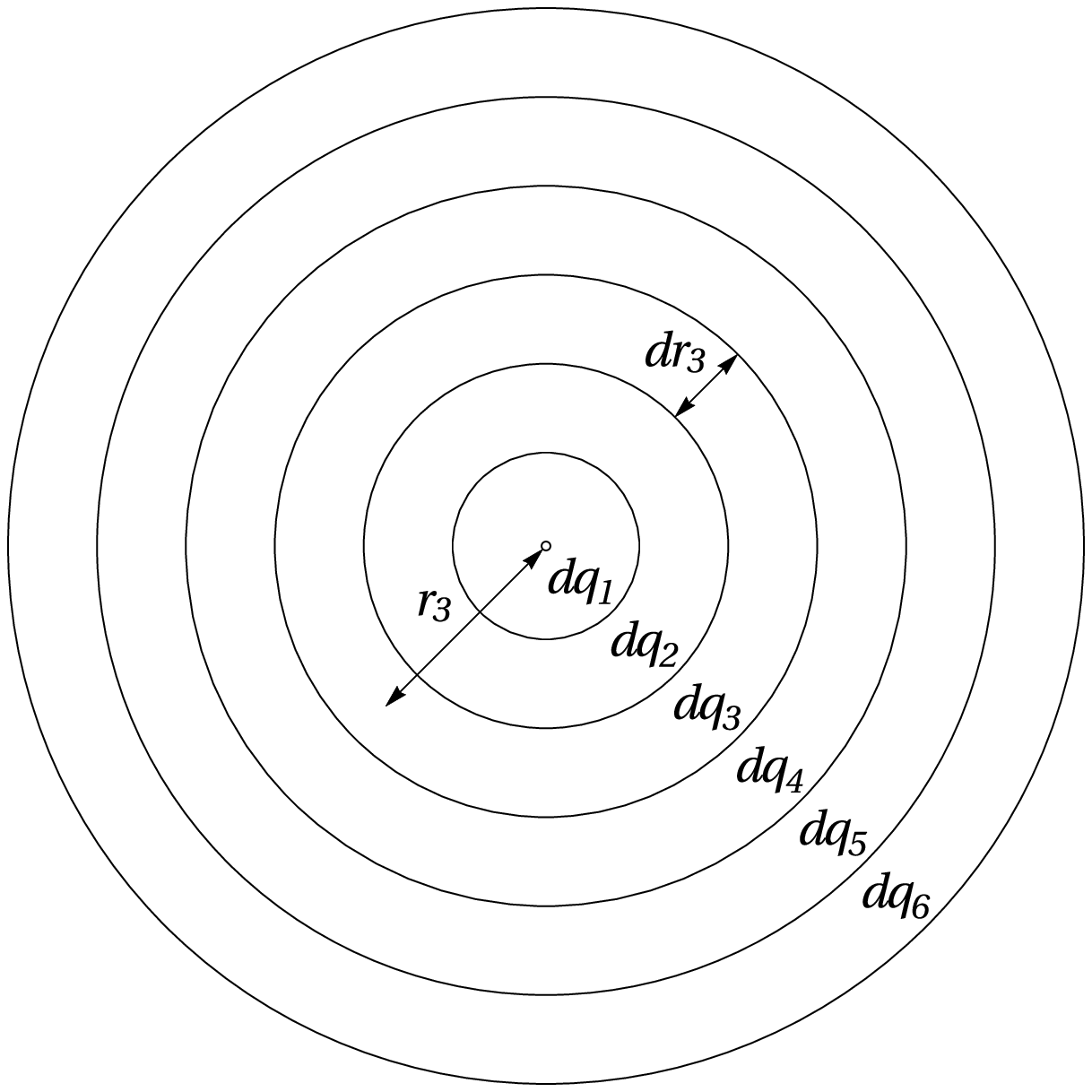}
\begin{picture}(0,0)
\put(-200,40){\includegraphics[scale=0.3]{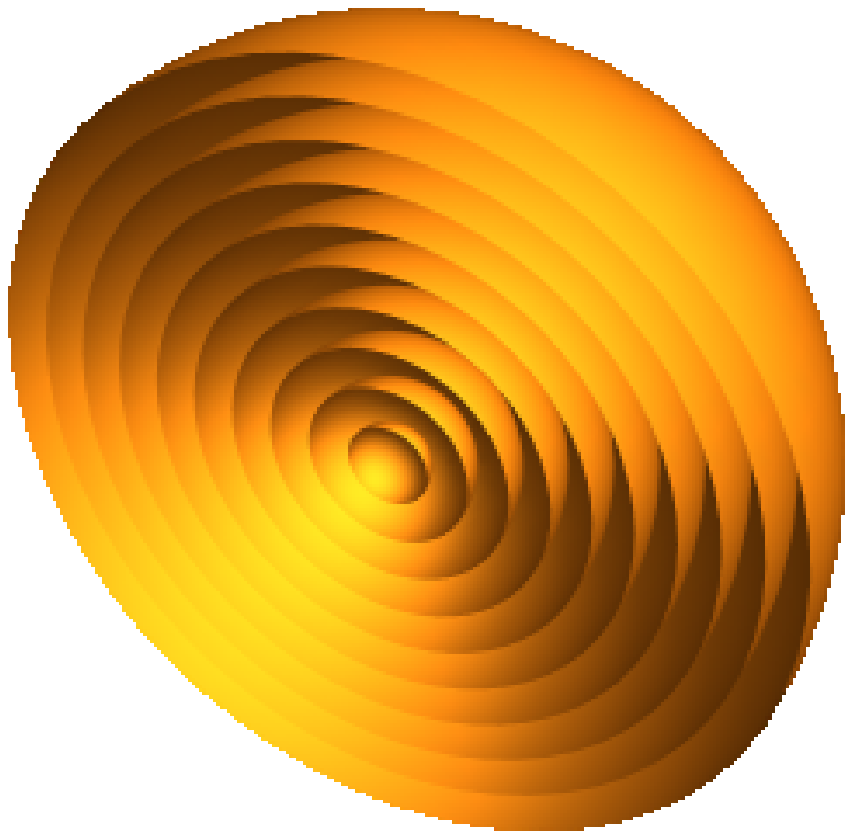}}
\end{picture}
\caption{\label{fig:shells}Schematic of the spherical distribution divided into spherical shells. The inset shows a 3-dimensional rendering of the distribution cut in half with the spherical shell boundaries.}
}
\end{figure}

Because we will construct our spherical distribution from spherical shells of charge, we will draw heavily from results for rigid spherical shells, which may be found in the references above. We follow a derivation that is similar to the development of the equations of motion for a rigid spherical shell in Ref.~\cite{yaghjian}, and when necessary, equations will be borrowed from there. The ``field reaction'' or ``radiation reaction'' (as well as the contribution of the self-electromagnetic field to the inertial mass of the distribution) comes from momentum transfer between the charge and its field via the self-electromagnetic force. Once we have solved for this, we will apply Newton's second law, accounting for all momentum transfer to and from the charge.

The derivation below may be summarized as: (1) calculate the self-electromagnetic force from the entire distribution on a constituent shell of charge (to zeroth order in the size of the distribution); (2) apply Newton's second law on the constituent spherical shell, including the self-electromagnetic force, and any necessary binding forces; (3) integrate this over the distribution to obtain the center-of-mass equations of motion; and (4) integrate this after subtracting the center-of-mass motion to obtain equations of motion for the net internal dipolar motion.

\section{Low-velocity equations of motion}
\subsection{Center-of-Mass Equation of Motion}
Assume our distribution, when in static equilibrium, is perfectly spherical. In this case, the distribution may be segregated into spherical shells. This is shown in Fig.~\ref{fig:shells}. Each shell contains charge $dq_i$ (where $i=1,2,3...N$ labels the shell), and has thickness $dr_i$. Taking the limit as $dr_i\rightarrow 0$, $N\rightarrow\infty$, all of the $i^{\rm th}$ shell's charge is uniformly distributed on the shell and concentrated at radius $r_i$; then $dq_i$ may be thought of as a continuous function of $r_i$, $dq_i\equiv dq(r_i)$. 

\begin{figure}
{
\includegraphics[scale=0.3]{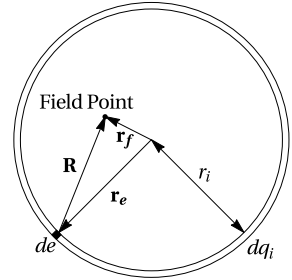}
\caption{\label{fig:singleShell}Schematic for calculation of electric field from a single shell, $dq_i$. $\vec{r}_e$ is the position of $de$, $\vec{r}_f$ is the position of the field point, and $\vec{R}=\vec{r}_f-\vec{r}_e$.}
}
\end{figure}

Now assume the shells accelerate (while maintaining their shape in their proper frame). The electric field at a field point, $\vec{r}_f$, due to a small (point-like) charge $de$ on an accelerating shell of charge that is spherical in its proper frame is\cite{yaghjian} 
\begin{equation}
\begin{array}{l}
d\vec{E} = \frac{1}{4 \pi \epsilon _0}\left\{\frac{\hat{R}}{R^2}+\left(\frac{1}{2 c^2 R}\left(\frac{\vec{r}_e\cdot \dot{\vec{u}}_i}{c^2}-1\right)\right)\left((\hat{R}\cdot \dot{\vec{u}}_i)\hat{R}+\dot{\vec{u}}_i\right)\right.\\
\left.+\frac{3}{8} \frac{\hat{R}}{c^4}\left(\left(\hat{R}\cdot \dot{\vec{u}}_i\right)^2-\dot{u}_i^2\right)+\frac{3 (\hat{R}\cdot \dot{\vec{u}}_i)\dot{\vec{u}}_i}{4 c^2}+\frac{2\ddot{\vec{u}}_i}{3 c^2}+O(R)\right\}de,
\end{array}
\end{equation}
where $\vec{r}_e$ is the position of the differential point of charge $de$, $\vec{R}=\vec{r}_f-\vec{r}_e$, $\hat{R}$ is the unit vector associated with $\vec{R}$, and $\vec{u}_i$ is the velocity the center of the shell, $dq_i$, as a function of time. See Fig.~\ref{fig:singleShell}. In this and all following equations, dots above variables signify time derivatives, bold variables are vectors, hatted variables are unit vectors, and unbolded italic variables are scalars. When an unbolded italic (hatted) variable has the same name as a bold variable, it is the magnitude (unit vector) of that vector. 

Integrating $d\vec{E}$ over shell $dq_i$ yields the electric field due to a rigid spherical shell in arbitrary motion to zeroth order in $R$:
\begin{widetext}
\begin{equation}
\begin{array}{c}
d\vec{E}_{r_i\geq r_f}=\frac{dq_i}{4 \pi  \epsilon _0}\left(\frac{2\ddot{\vec{u}}_i}{3 c^3}-\frac{2 \dot{\vec{u}}_i}{3 c^2 r_i}+\frac{4 \left( (\dot{\vec{u}}_i\cdot \vec{r}_f)\dot{\vec{u}}_i-\frac{1}{3} \dot{u}_i^2  \vec{r}_f \right)}{5 c^4 r_i}\right)\\
d\vec{E}_{r_i< r_f}=
\frac{dq_i}{4 \pi  \epsilon _0}\left(
\begin{array}{c}
\frac{\hat{r}_f}{r_f^2}+\frac{2\ddot{\vec{u}}_i}{3 c^3}-\do \left(\frac{r_i^2}{6 c^2 r_f^3}+\frac{1}{2 c^2 r_f}\right){\dot{\vec{u}}_i}+  \left(\frac{r_i^4}{20 c^4 r_f^4}+\frac{3}{4 c^4}\right) (\dot{\vec{u}}_i\cdot \hat{r}_f)\dot{\vec{u}}_i+\dot{u}_i^2 \left(\frac{r_i^4}{40 c^4 r_f^4}+\frac{r_i^2}{12 c^4 r_f^2}-\frac{3}{8 c^4}\right)\hat{r}_f \\
+ \left(\frac{r_i^2}{2 c^2 r_f^3}-\frac{1}{2 c^2 r_f}\right) (\dot{\vec{u}}_i\cdot \hat{r}_f)\hat{r}_f+ \left(-\frac{r_i^4}{8 c^4 r_f^4}-\frac{r_i^2}{4 c^4 r_f^2}+\frac{3}{8 c^4}\right) \left(\dot{\vec{u}}_i\cdot \hat{r}_f\right)^2\hat{r}_f
\end{array}
\right).
\end{array}
\label{eFromDe}
\end{equation}
\end{widetext}

Using this result, we can calculate the force from one spherical shell, $dq_i$, on another with radius $r_j$, $dq_j\equiv dq(r_j)$. Under the assumption of non-relativistic velocity, the force on a differential point of charge $de$ on shell $dq_j$ is just $d\vec{F}=\int d\vec{E}d e$. Using the electric field from Eq.~\ref{eFromDe}, most terms integrate to zero and the force from $dq_i$ on $dq_j$ is
\begin{equation}
d\vec{F}_{ij}=\frac{dq_i dq_j m r_0}{q^2}\left(\frac{1}{c}\ddot{\vec{u}}_i-\frac{1}{r_>}\dot{\vec{u}}_i\right)+O(R)
\label{F12}
\end{equation}
\begin{equation}
r_0\equiv\frac{q^2}{6 \pi \epsilon_0 m c^2},
\end{equation}
where $r_>$ is the greater of the radii of the spheres ($r_i$ and $r_j$), and $r_0$ is a constant with units of length, which will simplify results that follow. $m$ is the inertial mass of the total distribution (what one would measure in a laboratory by taking the ratio of the external force to the center-of-mass acceleration); $q\equiv\int dq_i$ is the integrated charge of the distribution. For an electron, $r_0\approx 1.88$~fm.

Integrating over all of the source shells ($dq_i$), the net force on shell $dq_j$, including its self-force ($r_i=r_j$), is
\begin{equation}
d\vec{F}_j=d\vec{F}_{e}+d\vec{F}_b+\frac{m r_0 d q_j}{q^2}\left(\int \frac{\ddot{\vec{u}}_i}{c}dq_i-\int \frac{\dot{\vec{u}}_i}{r_>}dq_i\right).
\label{dF2}
\end{equation}
Here and below, unless otherwise stated, integrals cover the whole distribution. I've included an arbitrary external force $d\vec{F}_{e}$ and the necessary binding force $d\vec{F}_b$ to keep $dq_j$ concentric (and from exploding). Now say $dq_j$ is allowed to displace from its equilibrium position. If the charge is stable against such displacements, the binding force must change to return the sphere to concentricity. Leaving $d\vec{F}_b$ as the equilibrium binding force, define a ``restoring force,'' $d\vec{F}_r$, as the change in the net force on $dq_j$ due to the displacement from equilibrium. Note that a deviation from concentricity will also change the self-electromagnetic force on $dq_j$, which is also included in $d\vec{F}_r$. Therefore, with the addition of $d\vec{F}_r$, Eq.~\ref{dF2} is fairly general, even allowing small displacements of the constituent spherical shells from concentricity.

At this point, it is convenient to define a parameter that quantifies the net displacement of the spherical shells from equilibrium, which will be related to the total dipole moment of the charge. Call $\vec{r}_{ci}$ the position of the center of spherical shell, $dq_i$, and define
\begin{equation}
\vec{r}_a\equiv \frac{1}{q}\int \left(\vec{r}_{c i}-\vec{r}\right) d q_i=\frac{1}{q}\int \vec{r}_{c i} dq_i -\vec{r},
\label{ra}
\end{equation}
where $q$ is the total charge of the distribution and $\vec{r}$ is the position of the center of mass of the distribution. The net dipole moment of the distribution about the origin may be written as $\vec{p}=q(\vec{r}+\vec{r}_a)$, and about the center of mass, $\vec{p}_{cm}=q\vec{r}_a$. Noting that the velocity of the center of the $i^{\rm th}$ shell is $\vec{u}_i=\dot{\vec{r}}_{c i}$, and taking successive time derivatives, Eq.~\ref{ra} yields
\begin{equation}
\begin{array}{c}
\vec{u}_a\equiv\frac{1}{q}\int \vec{u}_i dq_i -\vec{u}\\
\dot{\vec{u}}_a\equiv\frac{1}{q}\int \dot{\vec{u}}_i dq_i -\dot{\vec{u}}\\
\ddot{\vec{u}}_a\equiv\frac{1}{q}\int \ddot{\vec{u}}_i dq_i -\ddot{\vec{u}}.
\label{ua}
\end{array}
\end{equation}
At this point, no assumption has been made that the charge on each shell is constant, but Eqs.~\ref{ua} imply that any radial transfer of charge between shells does not change the total dipole moment. It should be noted that any time variation of the charge on each shell must also be slow compared the size of the charge divided by $c$, so that Eq.~\ref{F12} remains valid using the charge at the present time.

Evaluating the first integral in Eq.~\ref{dF2}
\begin{equation}
\begin{array}{lll}
d\vec{F}_j&=&d\vec{F}_{e}+d\vec{F}_b+d\vec{F}_r\\
&+&\frac{m r_0 dq_j}{q^2}\left(\frac{q}{c}(\ddot{\vec{u}}+\ddot{\vec{u}}_a)-\int \frac{\dot{\vec{u}}_i}{r_>}dq_i\right).
\end{array}
\label{dF22}
\end{equation}
Integrating both sides over all spherical shells, $dq_j$, yields the net force on the charge distribution:
\begin{equation}
\vec{F}=\vec{F}_{e}+\vec{F}_b+m r_0\left(\frac{1}{c}(\ddot{\vec{u}}+\ddot{\vec{u}}_a)-\frac{1}{q^2}\iint \frac{\dot{\vec{u}}_i}{r_>}dq_idq_j\right).
\label{FNet}
\end{equation}
where $\vec{F}_b=\int d\vec{F}_b+\int d\vec{F}_r$.

If the distribution is stable, the shells will oscillate around concentricity. As different spherical shells may oscillate with different phases, I transform to the frequency domain in order to more easily account for these phase differences. The following convention for the Fourier transform and transform of a product of functions is used:
\begin{equation}
\begin{array}{c}
f(\omega) = \mathcal{F}(f(t))=\frac{1}{\sqrt{2\pi}}\int_{-\infty}^{\infty}f(t)e^{-i\omega t}dt\\
\mathcal{F}(f(t)g(t))=\frac{1}{\sqrt{2\pi}}f(\omega)*g(\omega),
\label{fourier}
\end{array}
\end{equation}
where * denotes a convolution in $\omega$ space (a product of more than two functions produces factors of $1/\sqrt{2\pi}$ for each product). In the frequency domain, Eq. \ref{FNet} is then
\begin{equation}
\begin{array}{lll}
\vec{F}(\omega)&=&\vec{F}_{e}+\vec{F}_b+\frac{m r_0}{c}(\ddot{\vec{u}}+\ddot{\vec{u}}_a)\\
&-&\frac{m r_0}{2 \pi q^2}\iint \frac{\dot{\vec{u}}_i}{r_>}*(dq_i*dq_j),
\end{array}
\label{FOmega}
\end{equation}
where $\omega$ dependence on the right hand side is implied. Now assume the motion of the centers of the spheres can be described as
\begin{equation}
\vec{r}_{c i}(r_i,\omega)=\vec{r}(\omega)+f_q(r_i,\omega)\vec{r}_a(\omega),
\label{form}
\end{equation}
where, again, $r_i$ is the radius of shell $dq_i$, and $f_q$ is a scalar complex function that describes the relative amplitude and phase of the displacement of the shell with respect to the net dipolar motion. This is a statement that the displacement of an individual spherical shell, $\vec{r}_{c i}-\vec{r}$, lies along the net vector $\vec{r}_a$, is proportional to $\vec{r}_a$, and the relative phase/magnitude of the displacement of each shell to the net dipolar displacement only depends on $\omega$. This ``linearity'' is similar to linear models of dielectric materials, where the relative permittivity is a function of $\omega$. Note we implicitly ignore non-dipolar motion, i.e. flexing of the spheres and motion not in the direction of $\vec{r}_a$, which would result in higher order moments.

As an aside, $f_q$ is constrained by the definition of $\vec{r}_a$ in Eq.~\ref{ra}, and with some algebra, one finds
\begin{equation}
\frac{1}{q}\int f_q(r_i)dq_i=1.
\label{formNorm}
\end{equation}
The function $f_q$ can be viewed as describing the relative stiffness of parts of the distribution, and is set as a parameter of the model; Eq.~\ref{formNorm} sets the overall scale of $f_q$.

Using Eq.~\ref{form}, Eq. \ref{FOmega} becomes
\begin{equation}
\begin{array}{lll}
\vec{F}(\omega)&=&\vec{F}_{e}+\vec{F}_b+\frac{m r_0}{c}(\ddot{\vec{u}}+\ddot{\vec{u}}_a)\\
&-&\left(m k_0*\dot{\vec{u}}+\frac{m r_0}{2 \pi q^2}\iint \frac{f_q(r_i)\dot{\vec{u}}_a}{r_>}*(dq_i*dq_j)\right)
\end{array}
\label{fLastTerm}
\end{equation}
\begin{equation}
k_0\equiv\frac{r_0}{2\pi q^2}\iint\frac{dq_i*dq_j}{r_>}.
\label{k0}
\end{equation}
If the $dq$'s are constant in time, the convolutions can be performed noting they transform as $dq_{\rm const}(t)\rightarrow\sqrt{2\pi}dq_{\rm const}\delta(\omega)$. Therefore, in the case of constant $dq$, one obtains the net force on the distribution
\begin{equation}
\vec{F}(\omega)=\vec{F}_{e}+\vec{F}_b+\frac{m r_0}{c}(\ddot{\vec{u}}+\ddot{\vec{u}}_a)-m\left(k_0\dot{\vec{u}}+k_1\dot{\vec{u}}_a\right),
\label{FOmegaConst}
\end{equation}
where
\begin{equation}
k_1\equiv\frac{r_0}{q^2}\iint\frac{f_q(r_i)}{r_>}d q_j d q_i.
\label{k1}
\end{equation}
Identifying the force on the left hand side as the momentum imparted to the charge, the force must also be
\begin{equation}
\begin{array}{rcl}
\vec{F}(\omega)&=&\int \dot{{\vec{u}}}_j dm_{q j}=m_q \dot{\vec{u}}+\dot{\vec{u}}_a\int f_q(r_j) dm_{q j}\\
&=&m_q\dot{\vec{u}}+m k_{\rm ma}\dot{\vec{u}}_a,
\end{array}
\label{newton}
\end{equation}
where $dm_{q j}$ is the inertial mass inherent to the charge $dq_j$, $m_q\equiv\int dm_{q j}$, and $k_{\rm ma}\equiv \frac{1}{m}\int f_q(r_j) dm_{q j}$. Note that $m_q$ is not what is measured as the inertial mass of the distribution in the laboratory, since the inertial mass due to the charge's field is not included in $m_q$. This is discussed more below.

The binding force may be viewed as the momentum imparted to the charge by other matter attached to the charge distribution; this matter is responsible for binding the charge (canceling the self-electromagnetic forces) at equilibrium. For simplicity, I assume the other matter travels at the center-of-mass velocity throughout the distribution. Otherwise, we would need to track the dipolar motion of the other mass as well, which is beyond the scope of this paper.

Without knowing more of its nature, we can write the differential binding force on $dq_j$ as $d\vec{F}_b=-dm_{{\rm other},j}\dot{\vec{u}}$, where $dm_{{\rm other},j}\dot{\vec{u}}$ is the force by $dq_j$ on the ``other'' matter as the other matter accelerates; then apply Newton's third law. Note that $dm_{{\rm other},j}$ is not necessarily {\em located} on the shell $dq_j$; consider it the portion of the other mass over the {\em entire} distribution responsible for keeping $dq_j$ at its equilibrium position. Integrating, $\vec{F}_b=\int d\vec{F}_b=-m_{\rm other}\dot{\vec{u}}$. The function, $dm_{{\rm other},j}=dm_{\rm other}(r_j)$, must be set as a parameter of the model.

Implementing these relations, and rearranging Eq. \ref{FOmegaConst}
\begin{equation}
\begin{array}{lll}
(m_q + m k_0+m_{\rm other})\dot{\vec{u}}&=&\vec{F}_{e}+\frac{m r_0}{c}(\ddot{\vec{u}}+\ddot{\vec{u}}_a)\\
&-&m\left(k_{\rm ma}+k_1\right)\dot{\vec{u}}_a.
\end{array}
\label{uEq1}
\end{equation}
In the laboratory, the total inertial mass is measured (in the low velocity limit) by taking the ratio of $\vec{F}_e$ to $\dot{\vec{u}}$. Experimentally, of course, one would need to first remove the effects of the field reaction terms proportional to $\ddot{\vec{u}}$, and terms proportional to $\dot{\vec{u}}_a$ and $\ddot{\vec{u}}_a$. One may use this fact in Eq.~\ref{uEq1} to equate the coefficient in front of $\dot{\vec{u}}$ to the observed inertial mass $m$,
\begin{equation}
m=m_q + m k_0+m_{\rm other}.
\label{mq}
\end{equation}
Eq.~\ref{uEq1} then simplifies to
\begin{equation}
\dot{\vec{u}}=\frac{1}{m}\vec{F}_{e}+\frac{r_0}{c}(\ddot{\vec{u}}+\ddot{\vec{u}}_a)-\left(k_{\rm ma}+k_1\right)\dot{\vec{u}}_a.
\label{uEqFinal}
\end{equation}
This is the low-velocity equation of motion for the center of mass of a small spherical distribution of charge, if the charge per shell is constant in time. 

Another aside: Eq.~\ref{mq} acts as a constraint on the mass terms. Therefore, if $m_{\rm other}$ is set, then $m_q$ is automatically set by Eq.~\ref{mq}, or vice versa; any mass model that is chosen for $dm_{q}$ and $dm_{\rm other}$ must satisfy Eq.~\ref{mq} for consistency.

The constant, $k_0 m$, can be identified as the inertial mass of the self-electromagnetic field, as it is the coefficient of the acceleration that comes from the momentum transfer between the self-electromagnetic field and the charge. Relabeling Eq.~\ref{mq} for clarity, $m=m_q+m_{\rm field}+m_{\rm other}$.

\subsection{Dipolar Equation of Motion}
In order to fully determine the motion of the distribution, we need a second dynamic equation for $\vec{r}_a$. In order to do this, we need to make some assumption on the mass model used for the charge, i.e. $dm_{q j}$. Again, I note that $dm_{q j}$ is the inertial mass of $dq_j$ {\em without} its field. We assume one of two cases, (1) $dm_{q j}$ is a function of $dq_j$, and is non-zero for a shell of non-zero $dq_j$; (2) $dm_{q j}=0$. First, we treat case 1, and case 2 is discussed below. Under case 1, and imposing Newton's second law on $dq_j$, Eq.~\ref{dF22} is also equal to
\begin{equation}
d\vec{F}_j=\dot{\vec{u}}_j d m_{qj}.
\label{dqjN2}
\end{equation}
Replacing $d\vec{F}_j$ in Eq.~\ref{dF22}, subtracting both sides by $\dot{\vec{u}}dm_{q j}$, and multiplying by the factor $dq_j/dm_{q j}$ (assuming a well defined limit in the case of $dq_j\rightarrow 0$):
\begin{equation}
\begin{array}{lll}
(\dot{\vec{u}}_j-\dot{\vec{u}})dq_j&=&-\dot{\vec{u}}dq_j+\frac{dq_j}{dm_{qj}}[d\vec{F}_{e}+d\vec{F}_b+d\vec{F}_r\\
&+&\frac{m r_0 dq_j}{q^2}\left(\frac{q}{c}(\ddot{\vec{u}}+\ddot{\vec{u}}_a)-\int \frac{\dot{\vec{u}}_i}{r_>}dq_i\right)].
\end{array}
\label{uaEq0}
\end{equation}
Define $\rho_{qm}(r_j)\equiv\frac{dq_j}{dm_{q j}}$ as the charge density divided by its inertial mass density as a function of the radius of sphere $j$. Integrating over $dq_j$ gives:
\begin{equation}
\begin{array}{lll}
q\dot{\vec{u}}_a&=&-q\dot{\vec{u}}+\int\rho_{qm}(r_j)[d\vec{F}_{e}+d\vec{F}_b+d\vec{F}_r\\
&+&\frac{m r_0 dq_j}{q^2}\left(\frac{q}{c}(\ddot{\vec{u}}+\ddot{\vec{u}}_a)-\int \frac{\dot{\vec{u}}_i}{r_>}dq_i\right)].
\end{array}
\label{uaEq1}
\end{equation}
Transforming to the frequency domain, we have
\begin{equation}
\begin{array}{lll}
q\dot{\vec{u}}_a(\omega)&=&-q\dot{\vec{u}}+\int\frac{\rho_{qm}(r_j)}{\sqrt{2\pi}}*[d\vec{F}_{e}+d\vec{F}_b+d\vec{F}_r\\
&+&\frac{m r_0 dq_j}{\sqrt{2\pi}q^2}*\left(\frac{q}{c}(\ddot{\vec{u}}+\ddot{\vec{u}}_a)-\int \frac{\dot{\vec{u}}_i*dq_i}{\sqrt{2\pi}r_>}\right)].
\end{array}
\label{uaEq2}
\end{equation}
Assuming the $dq$'s are constant in time, using the fact that $d\vec{F}_b=-dm_{\rm other}\dot{\vec{u}}$, and using Eq.~\ref{form}, we obtain
\begin{equation}
\begin{array}{lll}
q\dot{\vec{u}}_a(\omega)&=&-q\dot{\vec{u}}+\int\rho_{qm}(r_j)(d\vec{F}_{e}+d\vec{F}_r)-q k_{b0}\dot{\vec{u}}\\
&+&k_{e0} \frac{q r_0}{c}(\ddot{\vec{u}}+\ddot{\vec{u}}_a)-q(k_{a0}\dot{\vec{u}}+k_{a1}\dot{\vec{u}}_a)
\end{array}
\label{uaEq3}
\end{equation}
\begin{eqnarray}
k_{b0}\equiv\frac{1}{q}\int\rho_{qm}(r_j)dm_{{\rm other},j}\\
k_{e0}\equiv\frac{m}{q^2}\int\rho_{qm}(r_j)dq_j\\
k_{a0}\equiv\frac{mr_0}{q^3}\iint\frac{\rho_{qm}(r_j)}{r_>}dq_idq_j\\
k_{a1}\equiv\frac{mr_0}{q^3}\iint\frac{\rho_{qm}(r_j)f_q(r_i)}{r_>}dq_idq_j.
\label{kValues}
\end{eqnarray}
Divide by $q$ and gather the $\dot{\vec{u}}_a$ terms,
\begin{equation}
\begin{array}{c}
(1+k_{a1})\dot{\vec{u}}_a(\omega)=\frac{1}{q}\int\rho_{qm}(r_j)(d\vec{F}_{e}+d\vec{F}_r)\\
-(1+k_{a0}+k_{b0})\dot{\vec{u}}+k_{e0} \frac{r_0}{c}(\ddot{\vec{u}}+\ddot{\vec{u}}_a).
\end{array}
\label{uaEq4}
\end{equation}
Now assume the displacement of each spherical shell is small, so the restoring force can be assumed linear in the displacement of the center of shell $j$ from the center of mass ($\vec{r}_{c j}-\vec{r}$):
\begin{equation}
d\vec{F}_r(r_j)=-k_{rr}(r_j)\frac{m c^2}{r_0^3}(\vec{r}_{c j}-\vec{r})dr_j,
\label{dFr}
\end{equation}
where $k_{rr}$ is a positive function of the radius; $k_{rr}$ describes the strength of the restoring force and is set as a model parameter. Eq.~\ref{dFr} is the first term in the power series expansion about equilibrium for an arbitrary smooth force that depends only on the position of the center of the shell. Evaluating the restoring integral,
\begin{equation}
\frac{1}{q}\int\rho_{qm}(r_j)d\vec{F}_r=-k_r\frac{c^2}{r_0^2}\vec{r}_a
\label{restoreInt}
\end{equation}
\begin{equation}
k_r\equiv \frac{m}{q r_0}\int\rho_{qm}(r_j)k_{rr}(r_j)f_q(r_j)dr_j.
\label{kr}
\end{equation}

To determine the dynamics of $\vec{r}_a$ in an external electric field, $\vec{E}$, set $d\vec{F}_e=dq_j\vec{E}$. This assumes that  $\vec{E}$ is constant over the displacements $\vec{r}_{c j}-\vec{r}$, and the source of $\vec{E}$ is outside the entire distribution. With these assumptions, the external force integral becomes
\begin{equation}
\frac{1}{q}\int\rho_{qm}(r_j)d\vec{F}_e=k_{e1}\frac{q}{m}\vec{E},
\label{extInt}
\end{equation}
where $k_{e1}=k_{e0}$. If the source of the electric field is within the outermost sphere, the treatment is still valid, but the integral truncates at the position where the source of the electric field is located, and $k_{e1}$ is no longer equal to $k_{e0}$. Inserting the external and restoring force, and rearranging Eq.~\ref{uaEq4}, we have
\begin{equation}
\dot{\vec{u}}_a(\omega)=k_{m1}\frac{q}{m}\vec{E}-k_{ar}\frac{c^2}{r_0^2}\vec{r}_a-k_{au}\dot{\vec{u}}+k_{m0}\frac{r_0}{c}(\ddot{\vec{u}}+\ddot{\vec{u}}_a)
\label{uaEqFinal}
\end{equation}
\begin{equation}
k_{m0}\equiv k_{e0}/\left(1+k_{a1}\right)
\label{km0}
\end{equation}
\begin{equation}
k_{m1}\equiv k_{e1}/\left(1+k_{a1}\right)
\label{km1}
\end{equation}
\begin{equation}
k_{ar}\equiv k_r/\left(1+k_{a1}\right)
\label{kar}
\end{equation}
\begin{equation}
k_{au}\equiv \left(1+k_{a0}+k_{b0}\right)/\left(1+k_{a1}\right).
\label{kau}
\end{equation}
Eqs.~\ref{uEqFinal} and \ref{uaEqFinal} are the low-velocity center-of-mass and dipolar equations motion of a small (zeroth order in $R$) spherically distributed charge in an electric field where the displacement of its spherical constituents from concentricity is small enough to assume a linear restoring force, and the response of the spherical constituents is linear with respect to the bulk dipolar motion.

Assuming $f_q$ is real (individual motions are completely in or out of phase with net dipolar motion about the center of mass), and its magnitude does not vary with frequency, then the same treatment may be performed in the time domain, and the equations maintain the same form as they do in the frequency domain, {\em without} the need of constraining the $dq$'s to be constant.

Depending on the choice of charge distribution ($dq_j=dq(r_j)$) and the choice of $dm_{{\rm other}}(r_j)$, Eq.~\ref{mq} may require the charge mass model to have $dm_{q j}>0$, $dm_{q j}<0$, or $dm_{q j}=0$ (i.e. the charge's mass is entirely due to its self-electromagnetic field). In the case of $dm_q=0$, $\rho_{qm}$ is ill defined. This special case may be treated in the following way: Eq.~\ref{dqjN2} becomes $dF_j=0$, and this may be inserted into Eq.~\ref{dF22}. The result is somewhat simpler: in the case of $dm_{q j}=0$, Eq.~\ref{uaEqFinal} remains the same, but the constants are simplified,
\begin{equation}
{\rm In~the~case:~}dm_{q j}=0
\label{m00}
\end{equation}
\begin{equation}
k_{m0}= 1/k_{1}
\label{km00}
\end{equation}
\begin{equation}
k_{m1}= 1/k_1
\label{km10}
\end{equation}
\begin{equation}
k_{ar}= k_{r0}/k_{1}
\label{kar0}
\end{equation}
\begin{equation}
k_{au}= \left(k_{0}+\frac{m_{\rm other}}{m}\right)/k_{1}
\label{kau0}
\end{equation}
\begin{equation}
k_{r0}\equiv \frac{1}{r_0}\int k_{rr}(r_j)f_q(r_j)dr_j.
\label{kr0}
\end{equation}

\section{Arbitrary velocity (relativistic) equations of motion}
We now turn to the development of arbitrary-velocity equations of motion. The treatment is similar, the only difference being the force from shell $i$ on shell $j$ is not $d\vec{F}_{ij}=\int\vec{E_i}dq_j$, due to the time it takes for light signals to cross the distribution (see Ref.~\cite{yaghjian} Appendix B). The new integral is complicated, but the result is similar to Eq.~\ref{F12}:
\begin{equation}
d\vec{F}_{ij}=\frac{dq_i dq_j m r_0}{c q^2}\left(\ddot{\vec{u}}_{pi,\parallel}+\frac{1}{\gamma_i}\ddot{\vec{u}}_{pi,\perp}-\frac{c}{r_>}\frac{d}{dt}(\gamma_i\dot{\vec{u}}_i)\right),
\label{F12Rel}
\end{equation}
where $\gamma_i=1/\sqrt{1-\frac{u_i^2}{c^2}}$, the subscript $p$ denotes the variable is evaluated in the proper frame of the shell, $\parallel$ means the portion of the vector parallel to the velocity $\vec{u}_i$, and $\perp$ means the portion perpendicular to $\vec{u}_i$. The proper parallel and perpendicular second derivative of the velocity relate to their inertial frame counterparts as (Ref. \cite{yaghjian} A.22)
\begin{equation}
\begin{array}{lll}
\ddot{\vec{u}}_{pi,\parallel}&=&\gamma_i^4(\ddot{\vec{u}}_{i,\parallel}+3\frac{\gamma_i^2}{c^2}(\vec{u}_i\cdot\dot{\vec{u}}_i)\dot{\vec{u}}_{i,\parallel})\\
\ddot{\vec{u}}_{pi,\perp}&=&\gamma_i^3(\ddot{\vec{u}}_{i,\perp}+3\frac{\gamma_i^2}{c^2}(\vec{u}_i\cdot\dot{\vec{u}}_i)\dot{\vec{u}}_{i,\perp}).
\end{array}
\label{parPerp}
\end{equation}
In order to proceed, assume the velocity of the center of each shell, $\vec{u}_i$, only deviates from the center-of-mass velocity, $\vec{u}$, by an amount that is always small with respect to c and $\vec{u}$. Thus, $\gamma_i\approx\gamma$, where $\gamma=1/\sqrt{1-\frac{u^2}{c^2}}$. Make a similar assumption of smallness of deviation for the acceleration, $\dot{\vec{u}}_i$, while allowing $\vec{u}$, $\dot{\vec{u}}$ to be arbitrarily large, and Eqs.~\ref{parPerp} are approximately
\begin{equation}
\begin{array}{lll}
\ddot{\vec{u}}_{pi,\parallel}&\approx&\gamma^4(\ddot{\vec{u}}_{i,\parallel}+3\frac{\gamma^2}{c^2}(\vec{u}\cdot\dot{\vec{u}})\dot{\vec{u}}_{i,\parallel})\\
\ddot{\vec{u}}_{pi,\perp}&\approx&\gamma^3(\ddot{\vec{u}}_{i,\perp}+3\frac{\gamma^2}{c^2}(\vec{u}\cdot\dot{\vec{u}})\dot{\vec{u}}_{i,\perp}),
\end{array}
\label{parPerp2}
\end{equation}
where $\parallel$ and $\perp$ are now with respect to $\vec{u}$. We may now proceed as above replacing Eq.~\ref{F12} with Eq.~\ref{F12Rel}.

Because of the extra products of functions of time in Eq.~\ref{parPerp2}, working in the frequency domain is cumbersome. Therefore, I restrict myself to the case where $f_q$ is real and constant as a function $\omega$ (while allowing for non-constant $dq_i$), and develop the equations directly in the time domain. 

Eq.~\ref{FNet} becomes
\begin{equation}
\begin{array}{lll}
\vec{F}&=&\vec{F}_{e}+\vec{F}_b+m r_0\left(\frac{\gamma^2}{c}(\ddot{\vec{u}}_{\perp}+\ddot{\vec{u}}_{a,\perp}+\gamma^2(\ddot{\vec{u}}_{\parallel}+\ddot{\vec{u}}_{a,\parallel}))\right.\\
&+&3\frac{\gamma^4}{c^3}(\vec{u}\cdot\dot{\vec{u}})(\dot{\vec{u}}_{\perp}+\dot{\vec{u}}_{a,\perp}+\gamma^2(\dot{\vec{u}}_{\parallel}+\dot{\vec{u}}_{a,\parallel}))\\
&-&\left.\frac{1}{q^2}\iint \frac{1}{r_>}\frac{d}{dt}(\gamma\vec{u}_i)d q_i d q_j\right).
\end{array}
\label{FNet2}
\end{equation}
Using Eq.~\ref{form}, evaluating the integral, and imposing Newton's second law as above yields the analog to Eq.~\ref{uEqFinal}:
\begin{equation}
\begin{array}{lll}
\frac{d}{dt}(\gamma \vec{u})&=&\frac{1}{m}\vec{F}_{e}-\left(k_{\rm ma}+k_1\right)\frac{d}{dt}(\gamma\vec{u}_a)\\
&+&r_0\left(\frac{\gamma^2}{c}(\ddot{\vec{u}}_{\perp}+\ddot{\vec{u}}_{a,\perp}+\gamma^2(\ddot{\vec{u}}_{\parallel}+\ddot{\vec{u}}_{a,\parallel}))\right.\\
&+&\left.3\frac{\gamma^4}{c^3}(\vec{u}\cdot\dot{\vec{u}})(\dot{\vec{u}}_{\perp}+\dot{\vec{u}}_{a,\perp}+\gamma^2(\dot{\vec{u}}_{\parallel}+\dot{\vec{u}}_{a,\parallel}))\right),
\end{array}
\label{uEqRel}
\end{equation}
which is the arbitrary-velocity equation of motion for the center of mass of the spherical distribution. The equation of motion for $\vec{u}_a$ follows exactly as before, noting the differences between Eq.~\ref{F12} and Eq.~\ref{F12Rel}:
\begin{equation}
\begin{array}{lll}
\frac{d}{dt}(\gamma\vec{u}_a)&=&k_{m1}\frac{q}{m}\vec{E}-k_{ar}\frac{c^2}{r_0^2}\vec{r}_a-k_{au}\frac{d}{dt}(\gamma\vec{u})\\
&+&k_{m0}r_0\left(\frac{\gamma^2}{c}(\ddot{\vec{u}}_{\perp}+\ddot{\vec{u}}_{a,\perp}+\gamma^2(\ddot{\vec{u}}_{\parallel}+\ddot{\vec{u}}_{a,\parallel}))\right.\\
&+&\left.3\frac{\gamma^4}{c^3}(\vec{u}\cdot\dot{\vec{u}})(\dot{\vec{u}}_{\perp}+\dot{\vec{u}}_{a,\perp}+\gamma^2(\dot{\vec{u}}_{\parallel}+\dot{\vec{u}}_{a,\parallel}))\right).
\end{array}
\label{uaEqRel}
\end{equation}
These equations may be made somewhat more familiar using the identity\cite{yaghjian}
\begin{equation}
\vec{v}_\perp+\gamma^2\vec{v}_\parallel=\vec{v}+\frac{\gamma^2}{c^2}(\vec{v}\cdot\vec{u})\vec{u},
\label{parPerpIdentity}
\end{equation}
where $\vec{v}$ is any vector. The equations of motion are then
\begin{equation}
\begin{array}{lll}
\frac{d}{dt}(\gamma \vec{u})&=&\frac{1}{m}\vec{F}_{e}-\left(k_{\rm ma}+k_1\right)\frac{d}{dt}(\gamma\vec{u}_a)\\
&+&r_0\left(\frac{\gamma^2}{c}\left\{\ddot{\vec{u}}+\ddot{\vec{u}}_{a}+\frac{\gamma^2}{c^2}\left[(\ddot{\vec{u}}+\ddot{\vec{u}}_a)\cdot\vec{u}\right]\vec{u}\right\}\right.\\
&+&\left.3\frac{\gamma^4}{c^3}(\vec{u}\cdot\dot{\vec{u}})\left\{\dot{\vec{u}}+\dot{\vec{u}}_{a}+\frac{\gamma^2}{c^2}\left[(\dot{\vec{u}}+\dot{\vec{u}}_{a})\cdot\vec{u}\right]\vec{u}\right\}\right),
\end{array}
\label{uEqFinalRel}
\end{equation}
\begin{equation}
\begin{array}{lll}
\frac{d}{dt}(\gamma\vec{u}_a)&=&k_{m1}\frac{q}{m}\vec{E}-k_{ar}\frac{c^2}{r_0^2}\vec{r}_a-k_{au}\frac{d}{dt}(\gamma\vec{u})\\
&+&k_{m0}r_0\left(\frac{\gamma^2}{c}\left\{\ddot{\vec{u}}+\ddot{\vec{u}}_{a}+\frac{\gamma^2}{c^2}\left[(\ddot{\vec{u}}+\ddot{\vec{u}}_{a})\cdot\vec{u}\right]\vec{u}\right\}\right.\\
&+&\left.3\frac{\gamma^4}{c^3}(\vec{u}\cdot\dot{\vec{u}})\left\{\dot{\vec{u}}+\dot{\vec{u}}_{a}+\frac{\gamma^2}{c^2}\left[(\dot{\vec{u}}+\dot{\vec{u}}_{a})\cdot\vec{u}\right]\vec{u}\right\}\right).
\end{array}
\label{uaEqFinalRel}
\end{equation}
Eq.~\ref{uaEqFinalRel} reduces to the usual Lorentz-Abraham equation in the case of zero $\vec{u}_a$\cite{rohrlich1997}:
\begin{equation}
\begin{array}{lll}
\frac{d}{dt}(\gamma \vec{u})&=&\frac{1}{m}\vec{F}_{e}+\frac{r_0\gamma^2}{c}\left\{\ddot{\vec{u}}+\frac{\gamma^2}{c^2}(\ddot{\vec{u}}\cdot\vec{u})\vec{u}\right.\\
&+&\left.3\frac{\gamma^2}{c^2}(\vec{u}\cdot\dot{\vec{u}})\left[\dot{\vec{u}}+\frac{\gamma^2}{c^2}(\dot{\vec{u}}\cdot\vec{u})\vec{u}\right]\right\}.
\end{array}
\label{uEqReduce}
\end{equation}

\section{Conclusion}
In summary, Eqs.~\ref{uEqFinal} and \ref{uaEqFinal} are the low-velocity equations of motion for a small spherical distribution of charge interacting with an external electric field, allowing for small co-linear dipolar motion within the distribution. The charge on each shell was assumed constant in time; however, if the internal dipolar motion is perfectly in phase or out of phase with the bulk dipolar motion, then the derivation is valid for varying charge on each shell (as long as the variation does not directly affect the dipole moment).

The relativistic generalizations (assuming the internal dipolar motion about the center of mass is non-relativistic) are given in Eqs.~\ref{uEqFinalRel} and \ref{uaEqFinalRel}; in this case, it is assumed the internal dipolar motion of each shell is perfectly in phase or out of phase with the bulk dipolar motion, while the charge on each shell is allowed to vary with time. 

In these equations, the self-force has only been calculated up to zeroth order in the size of the distribution. For this to be valid, terms first order or higher in the radius of the distribution must be negligible. First and second order radiation reaction terms in the size of a spherical shell have been derived recently in Refs.~\cite{galley2010,forgacs2012,galley2012}. The magnitude of these terms may be used to test whether the smallness approximation here is appropriate. Also, if higher order multipole moments of the motion contribute significantly, the equations derived here will be inadequate. 

The equations presented here may be useful in studying the motion of spherical structures, which are prone to internal dipolar motion, such as ions or atoms, where a positive central core is surrounded by their corresponding electron clouds. Studying the effect of the interaction between the internal and bulk dipole radiation reaction is made possible. This is all done in the classical regime, so of course, if quantum effects must be taken into account, the underlying field theory for these equations is inadequate.

This theory also allows for the study of different classical mass models of spherical charge distributions.


\section{Acknowledgments}
\input acknowledgement.tex   


\end{document}

%% file: author_list.tex
%
\affiliation{Colorado School of Mines, Golden, Colorado, USA}
\author{P.D.~Flammer} 
\email{pflammer@mines.edu}
\affiliation{Colorado School of Mines, Golden, Colorado, USA}
\vskip 0.25cm

%% file: acknowledgement.tex
%
The author gratefully acknowledges fruitful discussion with Travis Kopp while reviewing the contents of this paper.